\documentclass{PoS}
\usepackage{amsmath}
\newcommand{\tr}{\operatorname{Tr}}
\usepackage{graphicx}

\title{Fractional Charge and Confinement of Quarks}

\ShortTitle{Fractional Charge and Confinement of Quarks}

\author{\speaker{Philipp Scior}, \; Sam R. Edwards\\%
        Theoriezentrum, Institut f\"ur Kernphysik, TU Darmstadt, 64289 Darmstadt, Germany\\
        E-mail: \email{scior@theorie.ikp.physik.tu-darmstadt.de}}

\author{Lorenz von Smekal\\
         Theoriezentrum, Institut f\"ur Kernphysik, TU Darmstadt, 64289 Darmstadt, Germany\\Institut f\"ur Theoretische Physik, Justus-Liebig-Universit\"at Gie{\ss}en, 35392 Gie{\ss}en, Germany\\
       E-mail: \email{lorenz.smekal@physik.tu-darmstadt.de}}

\abstract{In quantum chromodynamics with static quarks the confinement-deconfinement phase transition is connected to the spontaneous breaking of the global $Z_3$ center symmetry. This symmetry is lost when one considers dynamical quarks. Owing to the fractional electric charge of quarks, we recover a global $Z_6$ center symmetry when QCD is regarded as a part of the Standard Model.
We present results from QCD-like theories extended by electromagnetic interactions and show that the weak coupling limit of the QED part of the model results in a center-like symmetry with disorder in the vacuum. This can be seen explicitly in a character expansion of the fermion determinant. Further, we show that corresponding center averages project the fermion determinant on N-ality zero and discuss whether the additional center symmetry can be used to eliminate the fermion sign problem in QCD with fundamental quarks.}

\FullConference{31st International Symposium on Lattice Field Theory LATTICE 2013\\
                 July 29 -- August 3, 2013\\
                 Mainz, Germany}

\begin{document}

\section{Introduction}
Understanding the deconfinement phase transition of QCD has been a challenging puzzle for a long time. One area where we think confinement is understood quite well is pure SU(N) Yang-Mills theory. Here, the relevant degrees of freedom for the deconfinement transition are SU(N) spin variables (Polyakov loops) and deconfinement is connected to the spontaneous breaking of a $Z_N$ center symmetry \cite{Greensite:2003bk,Svetitsky:1982gs}.\\
In full QCD with dynamical quarks the importance of the $Z_3$ center symmetry for deconfinement is less clear. Center symmetry is broken explicitly by the quarks in the fundamental representation of the gauge group and the deconfinement transition at vanishing chemical potential becomes a smooth crossover \cite{Borsanyi:2010bp, Bazavov:2010sb}.\\
All the above neglects the electric and weak charges of the quarks. Normally it is assumed that electroweak interactions only lead to perturbative corrections to the QCD phase diagram. Note, though, that the Standard Model exhibits a global $Z_6$ center symmetry combining the centers of all its gauge groups \cite{Baez:2009dj}. The physical realization of this symmetry might have a non-trivial impact on confinement and the phase structure of the Standard Model. Here we will present results from QCD-like theories extended by electromagnetic interactions and fractionally charged quarks.

\section{A Hidden Symmetry of the Standard Model}
The Standard Model has a global $Z_6$ center symmetry \cite{Baez:2009dj}. The six center elements that act trivially on all fields are given by the combined transformations
\begin{equation}
 (e^{i 2 \pi /3} ,-1,e^{i \pi Y})\in \text{SU(3)}\times \text{SU(2)}\times \text{U(1)}_Y \; , 
\end{equation}
where $Y$ is the weak hypercharge operator. The $Z_6$ center group plays an important role in anomaly cancellation \cite{Geng:2000pf}. This symmetry is also necessary for the embedding the Standard Model's gauge group in a larger gauge group of a simple Grand Unified Theory, like for example SU(5) in the Georgi-Glashow model \cite{Georgi:1974sy}.\\
Electric charge $Q$ is related to the third component of the weak isospin $e^{i 2 \pi t_3}=-1 \in$ SU(2) and the hypercharge $Y$ via $Q=t_3 + Y/2$. By neglecting weak interactions and since the quarks carry fractional electric charges $Q=\frac{2}{3}e$ and $-\frac{1}{3}e$ we arrive at a  model with a $Z_3$ center, where the center elements generated by the combined transformations
\begin{equation}
(1,1),(e^{i 2\pi/3},e^{i 2\pi Q/e}),(e^{i 4\pi/3}, e^{i 4\pi Q/e}) \in \text{SU(3)}\times \text{U(1)}_{em} \; ,
\end{equation}
act trivially on all fields involved in the model. The presence of a global symmetry allows for its spontaneous breaking connected to a phase transition in the model. In the present case the breaking of the global $Z_3$ center symmetry in the presence of dynamical quarks could lead to an well-defined deconfinement phase transition. \newpage
\section{Two-Color Toy Model}
\begin{figure}
\begin{center}
 \includegraphics[width=.45 \textwidth]{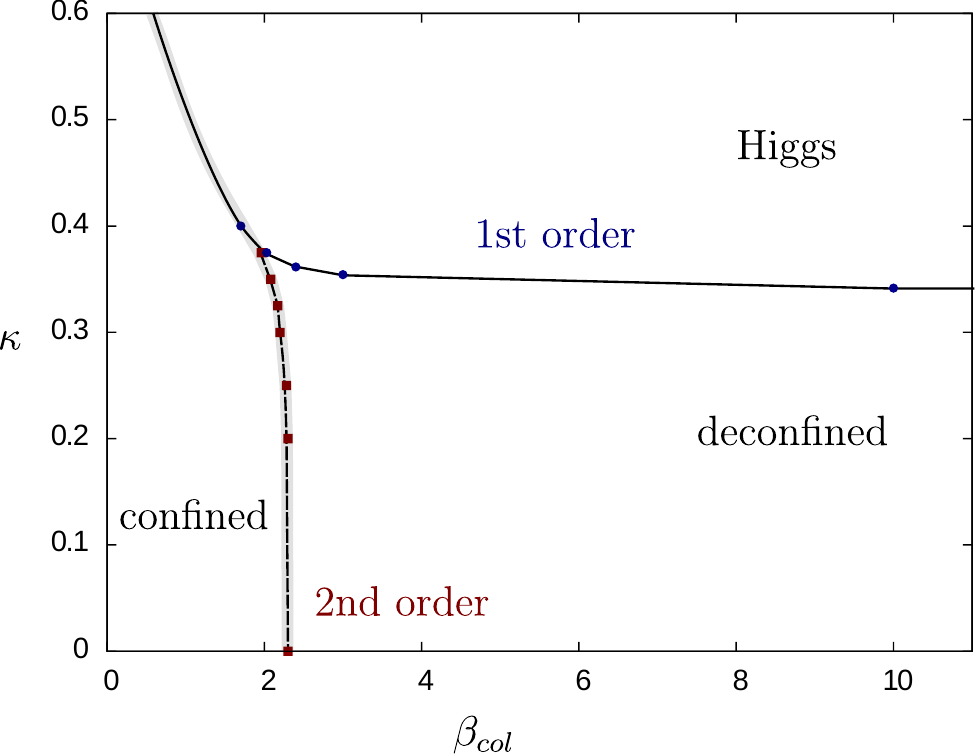} \hspace*{0.5 cm}
  \includegraphics[width=.49 \textwidth]{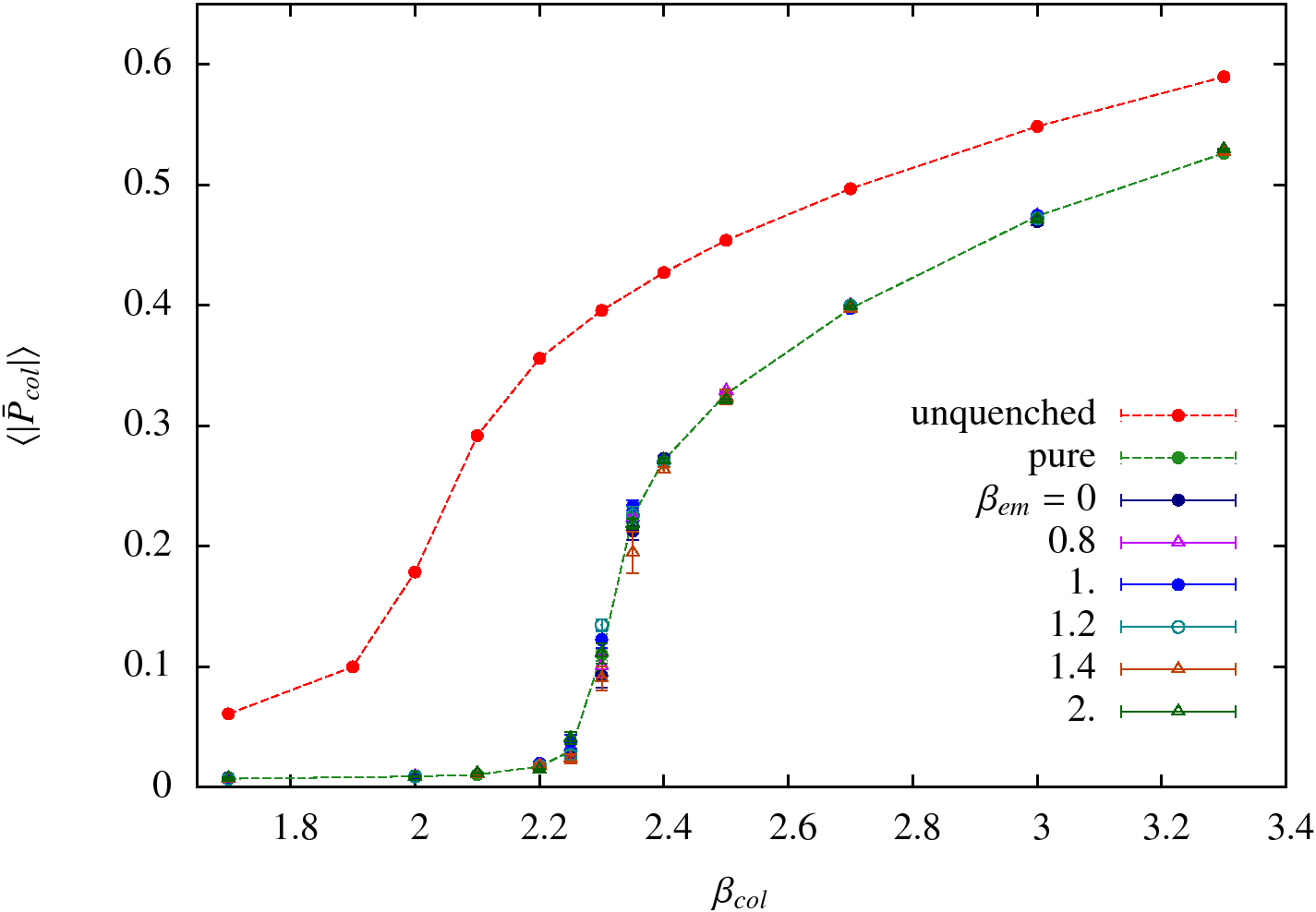}
  \caption{\textbf{left:} Phase diagram of the $\mathcal{O}(\kappa^4)$ effective action for $N_f=2$ fractionally charged Wilson quarks, for $N_t=4$. The center symmetry-breaking $\beta_{col}$ transition meets with a first order bulk transition that separates the phases in the low $\kappa$ region from a totally ordered Higgs-like phase. \textbf{right:} Volume averaged SU(2) Polyakov loop on $4 \times 16^3$ lattices, in the pure gauge gauge theory (green), 2-color QCD with $\kappa=0.15$ Wilson quarks (red), and the SU(2)$\times$U(1) toy model with fractionally charged quarks. The electromagnetic coupling $\beta_{em}$ here is varied from total U(1) disorder, $\beta_{em}=0$, to deep in the Coulomb phase, $\beta_{em}=2$.} 
\label{phases}
\end{center}
\end{figure}

For simplicity we considered a two-color QCD toy model for QCD with electromagnetic interactions and two flavors of fractionally charged Wilson quarks \cite{Edwards:2012tr}. To end up with a $Z_2$ center-like symmetry we choose $\pm \frac{1}{2}e$ as fractional charges for the quarks. Our action reads
\begin{equation}
 S=-\sum_p \left(\frac{\beta_{col}}{2} \tr[U_p]+ \beta_{em} \cos 2\varphi_p \right) + S_{f,W} \; , \label{action}
\end{equation}
where $S_{f,W}$ is the usual Wilson fermion action, with the distinction that parallel transporters for the quarks are products of an SU(2) color matrix and a U(1) phase, of the form
\begin{equation}
 U_\mu(x) e^{i \varphi_\mu(x)}, \hspace{.5cm}  U_\mu(x) \in \text{SU(2),} \hspace{.5cm}\varphi_\mu(x)\in [0,2\pi) \; . \label{link}
\end{equation}
The plaquettes are formed from the link variables $U_\mu$ and $\varphi_\mu$ in the usual way while the fractional charge is incorporated by the fact that the U(1) plaquette angle $2 \varphi_p$ is twice as large relative to the angle appearing in $U_\mu e^{i \varphi_\mu(x)}$. That is, a U(1) link $e^{i \varphi_\mu}=-1$ for quarks contributes as $e^{i 2\varphi_\mu}=+1$  to the gauge action.\\
It is apparent from \eqref{link} that the combination of the SU(2) center element $-1$ combined with the U(1) element $e^{i \varphi_\mu}=-1$ act trivially on the quarks. Therefore our theory has an explicit global $Z_2 = \{(-1,-1),(1,1) \}$ center symmetry combining the centers of the color and U(1) gauge groups, even in the presence of dynamical quarks.\\
As the electromagnetic gauge coupling $\beta_{em}$ is increased, the gauge action orders the U(1) links towards $e^{i \varphi_\mu}=\pm 1$, up to gauge transformations. The U(1) links are then essentially constrained to $Z_2$ links and the explicit breaking of the center symmetry by the fundamental quarks is undone via their coupling to an additional source of $Z_2 \subset$ U(1) topological disorder. For smaller U(1) gauge couplings, like i.e. in the confining phase of compact QED in 3+1 dimensions, the restoration of the $Z_2$ center symmetry is analogous to placing an Ising model in a fluctuating external magnetic field, or the Pecci-Quinn mechanism where the CP violating $\theta$ term of QCD is coupled to an axion field \cite{Peccei:1977ur}. In this case the symmetry breaking terms are suppressed by an additional source of disorder. The results of our simulations, displayed in Figure~\ref{phases} (right), show that the SU(2) Polyakov loop of our model is in agreement with the second order phase transition of pure SU(2) gauge theory, even in the vicinity of the bulk phase transition of the compact U(1) at $\beta_{em}\approx 1.01$. The behavior of the SU(2) Polyakov loop is in fact independent of the chosen coupling for the electromagnetic interaction $\beta_{em}$.\\
We have also explored the $(\beta_{col},\kappa)$ phase diagram of our theory using $\mathcal{O}(\kappa^4)$ hopping expansion and $N_t=4$ with the action
\begin{equation}
S_{eff}=-\sum_p \left(\frac{\beta_{col}}{2} \tr[U_p]+ \beta_{em} \cos 2\varphi_p \right) - 16 \kappa^4 \left(\sum_p \cos \varphi_p \cdot \tr[U_p] +  \sum_{x} \text{Re} \, P_{em} \cdot P_{col} \right) \; , \label{hopping}
\end{equation}
where $P_{em}$ and $P_{col}$ are the U(1)-, respectively SU(2)-Polyakov loops. Figure~\ref{phases} (left) shows the phase diagram in the limit $\beta_{em} \rightarrow \infty$. As $\kappa$ is increased from zero the deconfinement transition shifts towards smaller values of $\beta_{col}$ without changing its qualitative nature. When $\kappa$ is increased further, the transition sharpens and becomes first order, as seen in both  the U(1) and the SU(2) order parameter. This behaviour can be understood by looking at the action \eqref{hopping} and taking the combined limit $\beta_{em},\kappa\rightarrow~\infty$. This forces the SU(2) plaquette to take values $\pm 1$ according to the $Z_2 \subset$ SU(2) links. For large $\kappa$ the transition line must therefore terminate with the first order bulk transition of $Z_2$ gauge theory at $\beta_{col}=\ln(1+\sqrt{2})/2\approx 0.44$ \cite{Wegner:1984qt}. For $\beta_{col}>2.3$ center symmetry in the finite temporal direction is broken for all values of $\kappa$. Yet the order parameters exhibit an additional first order transition at $\kappa \approx 0.35$ where spacial center symmetry is broken too. The phase diagram of the model with the $\mathcal{O}(\kappa^4)$ effective action does not exhibit an analytic path between the different phases, see Figure \ref{phases} (left). In contrast, the existence of such an analytic path is ensured in an SU(2) model with fundamental Higgs field by the Fradkin-Shenker theorem \cite{Fradkin:1978dv}. The Fradkin-Shenker theorem is circumvented by our additional global $Z_2$ center symmetry. 

\section{Character Analysis of the Fermionic Weight}
The introduction of matter fields does not always break center symmetry explicitly, e.g. quarks in the adjoint representation of the gauge group respect center symmetry. That is different representations of fields transform differently under center transformations. Representations that respect center symmetry are called N-ality zero \cite{Greensite:2003bk}. In order to understand how a partition function with quarks in the fundamental representation of the SU(N) gauge group, coupled to an additional Z$_N$ field, is center symmetric, we restructure the partition function in such a way that the fermionic part is isolated for each gauge configuration
\begin{align}
Z&= \int [dU]  \exp[-S_{YM}] \int [d\psi, d \bar \psi] \sum_{[z \in Z_N]} \exp [-S_{f}] \; .
\end{align}
Using character analysis, it is possible to check which fermionic representations give a net contribution to the center averaged fermionic weight $Z_{f}=\int [d\psi, d \bar \psi] \sum_{[z \in Z_N]} \exp [-S_{f}]$.\\
Therefore one expands $Z_{f}$ into group characters $\chi_r(U)=\tr[M^{(r)}[U]]$ of the representations $r$
\begin{equation}
Z_{f}= \sum_r f_r \chi_r(u) \; ,
\end{equation}
where the sum runs over all irreducible unitary representations of the group \cite{Montvay:1994cy} and
\begin{equation}
f_r= \int [dU] \overline{\chi_r}(U) Z_{f} \; . \label{expansion}
\end{equation}
We are not able to solve all the integrals necessary to compute the expansion coefficients $f_r$ for any gauge group. Still we can use symmetry arguments in SU(2) and SU(3) to show that only N-ality zero representations are contributing to $Z_{f}$. If one now investigates the link-wise contribution to $f_r$
by using suitable parametrizations for the group elements and character, see  \cite{Murnaghan:1962, Reshnikoff:1966} for SU(3), one is able to show that for every link the integral over the class angles of the group gives zero except for those representations with N-ality zero. Therefore the character expansion of the Z$_3$ averaged $Z_{f}$ does not contain any parts with non-zero N-ality and does not break center symmetry explicitly.
\section{Implications for the Sign Problem}
The sign problem of QCD makes it difficult to explore the QCD phase diagram at finite chemical potential \cite{deForcrand:2010ys}. Is is common to seek insight from QCD-like theories that do not suffer from a sign problem such as e.g. two-color QCD, G$_2$-QCD or QCD with adjoint quarks \cite{Maas:2012wr, vonSmekal:2012vx}. Another peculiar point is that there is no center symmetry in G$_2$ or symmetry breaking by matter fields in adjoint QCD. Now the question arises, if we can remove the sign problem by projecting the fermionic weight of QCD on N-ality zero. From random matrix theory, we know that the complex representations of a group (i.e. the fundamental representation of SU(3)) lead to sign problems \cite{vonSmekal:2012vx, Dyson}. We can again use \eqref{expansion} to check, if the $Z_3$ averaging removes all complex representations from the partition function. Unfortunately this is not the case, since not all N-ality zero representations are strictly real, e.g.. the representation of SU(3) with the canonic labels (4,1) has N-ality zero but is complex. However Z$_3$ averaging removes a large fraction of the complex representations from the expansion \eqref{expansion}. Have we made the sign problem less severe by averaging over the center elements?\\
To test our hypothesis we have calculated the sign of the fermion determinant in Monte Carlo simulations of a  temporal one-link model with an artificial 4d Dirac structure as well as a Polyakov loop model. In both cases we employ the quenched approximation. \newline The YM action and Dirac operator for the one-link model read
\begin{figure}
\begin{center}
 \includegraphics[width=.48 \textwidth]{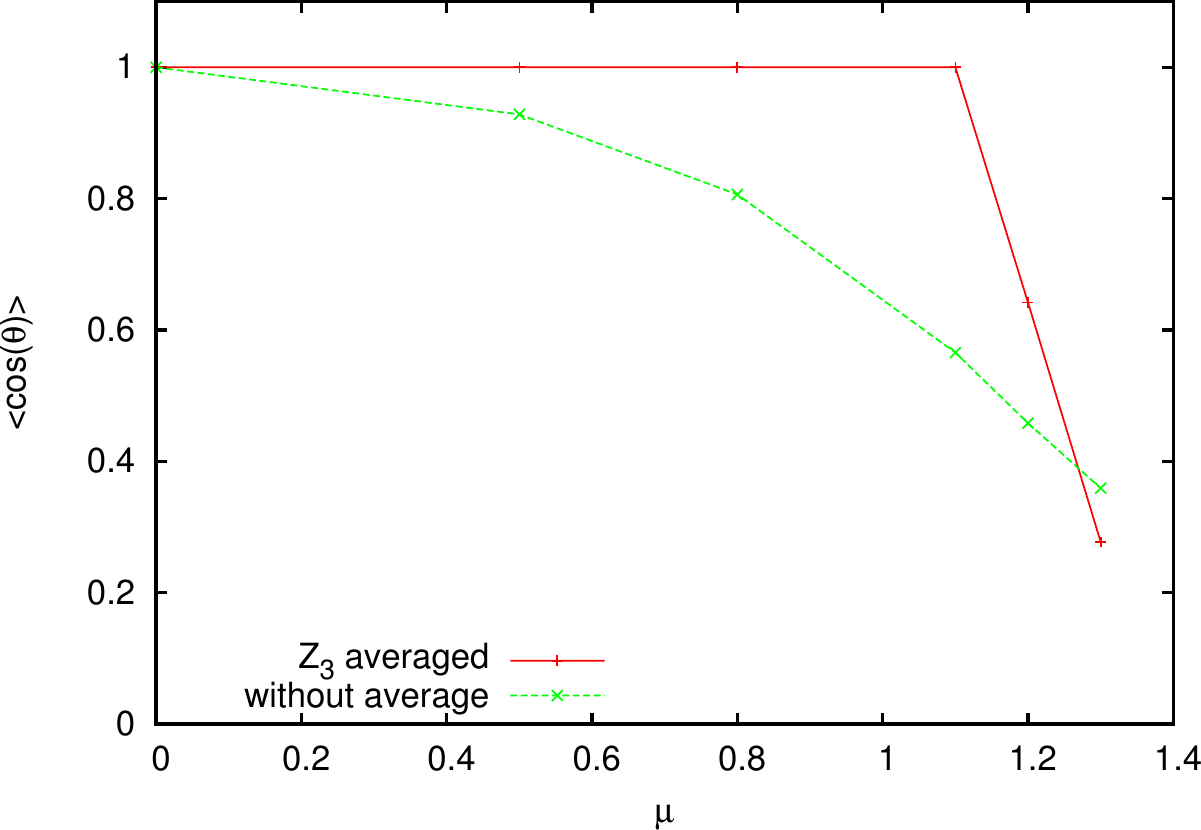} \hspace*{0.1 cm}
  \includegraphics[width=.48 \textwidth]{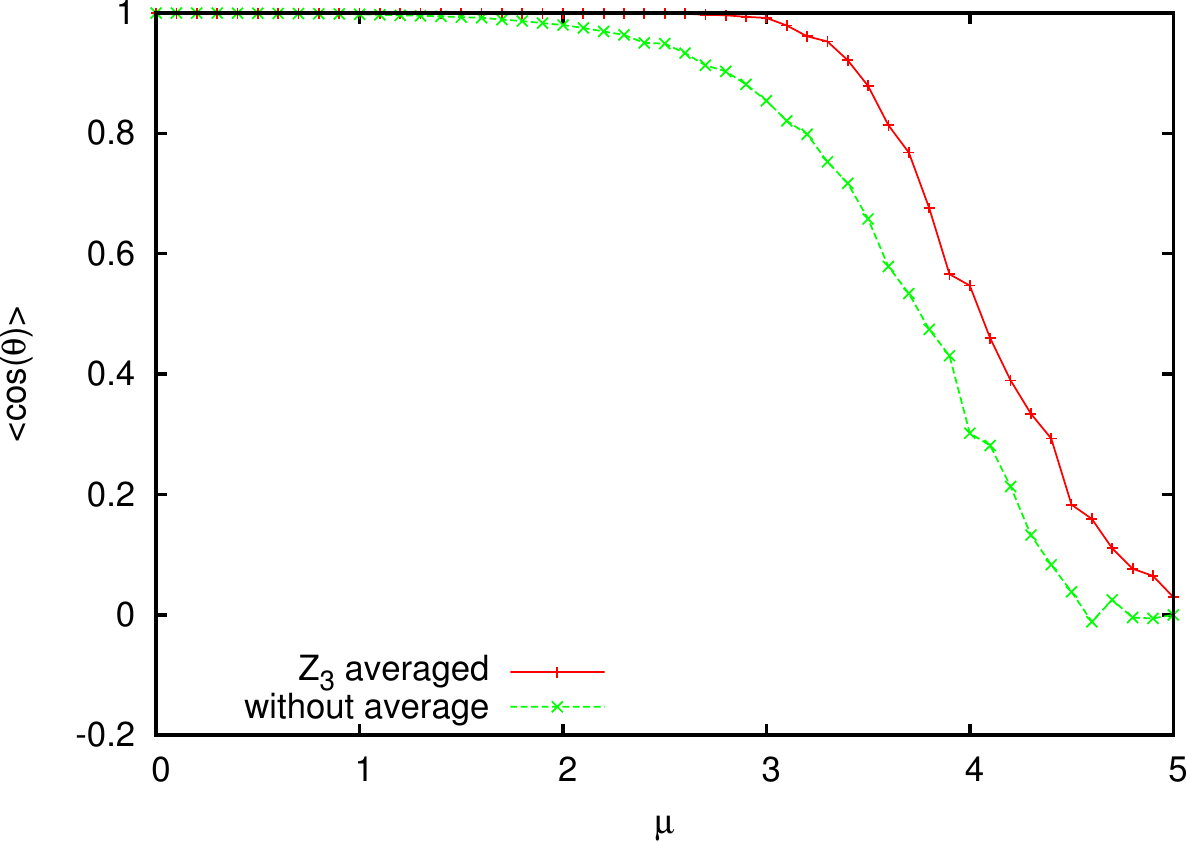} 
  \caption{\textbf{left:} Sign of the fermion determinants of a SU(3) one-link model, with (red) and without (green) Z$_3$ averaging of the fermion determinant. $\beta=0.66$, $m=1$  \textbf{right:} Sign of the fermion determinants of a SU(3) Polyakov loop model, with (red) and without (green) Z$_3$ averaging of the fermion determinant. $\lambda=0.1$, $\tilde{h}=0.0002$ $N_s=24$ and we set  $T=1$. }
\label{signs}
\end{center}
\end{figure}
\begin{align}
S_{YM}&= -\frac{\beta}{3} \text{Re}\tr[U] \; ,  \notag \\
D&=(m+4)-\frac{1}{2}\left( (1-\gamma_4) U e^\mu + (1+\gamma_4) U^\dagger e^{-\mu} \right) \; .
\end{align} 
Figure~\ref{signs} (left) shows the ensemble averages of the signs of the fermion determinant $cos(\theta)=\text{Re}\frac{\det D(\mu)}{|\det D(\mu)|}$ of a SU(3) one-link model and of the Z$_3$ averaged fermion determinant $\sum_z \det(zU)$. The sign of the Z$_3$ averaged fermion determinant stays close to one until it starts to fall off at about $\mu=1.2$. This result is similar to the large N results of Bringoltz \cite{Bringoltz:2010iy} in 1+1 dimensions and recent findings by Bloch et al. \cite{Bloch:2013ara} in 0+1 dimensional QCD with center averaging. However the authors in \cite{Bloch:2013ara} find that the fermion determinant remains exactly 1 for arbitrary values of $\mu$. The main difference in our results seem to be generated by the artificial 3+1 dimensional Dirac structure in our one-link model. \\
We also compared the signs of a center averaged and a non averaged fermion determinant in  Monte Carlo simulations of a quenched SU(3) Polyakov loop model. The effective YM action $S_{eff}$ and fermion determinant for one flavor $Q$ read
\begin{equation}
S_{eff}=- \lambda \sum_{<ij>} \text{Re} L_i L_j^* \; ,  \hspace*{0.5 cm} Q=\exp \left(- \sum_i h L_i + \overline{h} L^*_i \right)\; ,
\end{equation}
where the sum $<ij>$ runs over all pair of nearest neighbours. The couplings are defined by $h=\tilde{h} \,e^{\mu/T}$ and $\overline{h}(\mu)=h(-\mu)$. The results are shown in Figure~\ref{signs} (right). In this case, the difference between the signs of the determinant with and without center averaging is still qualitatively the same as in the one link model. However the effect of the center averaging is not as dramatic.

\section{Discussion} 
Given the fractional electric charges of quarks and the Z$_6$ center symmetry of the Standard model, it might not be accurate to study the effects of QCD without the electro-weak gauge groups. Our two-color simulations with electromagnetic interactions and fractionally charged quarks show the restoration of the second order deconfinement transition in the presence of dynamical quarks. In our model the fermion determinant is projected onto N-ality zero and so no longer breaks center symmetry explicitly.\\
Finally we showed that the introduction of fractionally charged quarks removed many, but not all complex representations of SU(3) from the fermionic weight. We discussed the implications for the fermion sign problem of QCD using an SU(3) one-link model and an SU(3) Polyakov loop model as qualitative examples. In both cases the onset of a sign problem was shifted to higher $\mu$. As a next step, it would be interesting to compare unquenched Polyakov loop simulations with $Z_3$ center average to Worm algorithm results of the non averaged theory to further investigate the influence of $Z_3$ averaging on the sign problem.

\paragraph*{Acknowledgements:}
The authors thank Jeff Greensite and Kurt Langfeld for helpful discussions and work on related subjects. This work was supported by the Helmholtz International Center for FAIR within the LOEWE initiative of the State of Hesse and the European Commission, FP7-PEOPLE-2009-RG No. 249203. Simulations were performed on the LOEWE-CSC High Performance Cluster at Goethe-Universit\"at Frankfurt am Main.

\end{document}